# COCCIALAB

*To discover the causes of social, economic and technological change*

**How sustainable environments have reduced the diffusion of coronavirus disease 2019: the interaction between spread of COVID-19 infection, polluting industrialization, wind (renewable) energy**

Mario COCCIA

CNR -- National Research Council of Italy

# How sustainable environments have reduced the diffusion of coronavirus disease 2019: the interaction between spread of COVID-19 infection, polluting industrialization, wind (renewable) energy

*Mario Coccia*
CNR -- National Research Council of Italy
Research Institute on Sustainable Economic Growth
Collegio Carlo Alberto, Via Real Collegio, 30-10024 Moncalieri (Torino, Italy)

*E*-mail: mario.coccia@cnr.it

Mario Coccia ORCID: http://orcid.org/0000-0003-1957-6731

## Abstract

The pandemic of coronavirus disease 2019 (COVID-19) is rapidly spreading all over the world, generating a high number of total deaths. The contemporary environmental and sustainability debate generates new or relatively unexplored problems. One of the current questions is: how high polluting industrialization and a less cleaner production affect the diffusion of COVID-19 infection? This study endeavors to explain the relation between air pollution and particulate compounds emissions, wind resources and energy, and the diffusion of COVID-19 infection to provide insights of sustainable policy to prevent future epidemics. The statistical analysis here focuses on case study of Italy, one of the countries to experience a rapid increase in confirmed cases and deaths. Results reveal two main findings: 1) cities in regions with high wind speed and a high wind energy production in MW have a lower number of infected individuals of COVID-19 infection and total deaths; 2) cities located in hinterland zones (mostly those bordering large urban conurbations) with high polluting industrialization, low wind speed and less cleaner production have a greater number of infected individuals and total deaths. Hence, cities with pollution industrialization and low renewable energy have also to consider low wind speed and other climatological factors that can increase stagnation of the air in the atmosphere with potential problems for public health in the presence of viral agents. Results here suggest that current pandemic of Coronavirus disease and future epidemics similar to COVID-19 infection cannot be solved only with research and practice of medicine, immunology and microbiology but also with a proactive strategy directed to interventions for a sustainable development. Overall, then, this study has to conclude that a strategy to prevent future epidemics similar to COVID-19 infection must also be based on sustainability science to support a higher level of renewable energy and cleaner production to reduce polluting industrialization and, as result, the factors determining the spread of coronavirus disease and other infections in society.

**Keywords**: Air Pollution, particulate compounds, Wind Energy, Wind Resource, Renewable Energy, Ventilation Potential, COVID-19, Coronavirus Disease, Coronavirus Infection; SARS-CoV-2, Sustainable Growth, Cleaner Production

**JEL codes:** K32, Q01, Q20; Q40, Q50, Q53, Q54, Q55, Q56, Q58, O13, O32, O33

# INTRODUCTION

The contemporary environmental and sustainability debate has new or relatively unexplored topics that continually emerge in science. This study provides an investigation for the exploration of causes, consequences and sustainable policy responses linked to diffusion of Coronavirus disease 2019 in a context of environmental and sustainability science.

The Coronavirus disease 2019 (COVID-19) produces minor symptoms in most people, but is also the cause of severe respiratory disorders and death of many individuals worldwide (Ogen, 2020; Dantas et al., 2020). The Coronavirus infection, started in China in 2019, is an on-going global health problem that is generating a socioeconomic crisis and negative world economic outlook projections (Saadat et al., 2020). Manifold studies suggest a possible relation between air pollution and particulate compounds emissions and diffusion of COVID-19 infection (Fattorini and Regoli, 2020; Frontera et al., 2020). Scholars also state that high level of air pollution can increase viral infectivity and lethality of COVID-19 infection (Contini and Costabile, 2020). Conticini et al. (2020) argue that population living in regions with high levels of particulate compounds emissions has also a high probability to develop respiratory disorders because of infective agents. In fact, the highest level of COVID-19 infection is in the USA, Spain, Italy, UK, Russia, China, Brazil, France, etc. that are countries having in some regions a very high level of air pollution (Coccia, 2020; Frontera et al, 2020). Studies confirm correlations between exposure to air pollution, diffusion and virulence of the SARS-CoV-2 within regions with population having a high incidence of respiratory disorders, such as chronic obstructive pulmonary disease (COPD) and Lung Cancer (Fattorini and Regoli, 2020; Coccia, 2014, 2015). Lewtas (2007) shows that exposures to combustion emissions and ambient fine particulate air pollution are associated with genetic damages. Long-term epidemiologic studies report an increased risk of all causes of mortality, cardiopulmonary mortality, and lung cancer mortality associated with increasing exposures to air pollution (cf., Coccia and Wang, 2015). Ogen (2020, p.4) finds that high $NO_2$ concentrations associated with downwards airflows cause of $NO_2$ buildup close to the surface. This geographical aspect of regions, associated with specific atmospheric conditions of low wind, prevents the dispersion of air pollutants, which are one of the factors of a high incidence of respiratory disorders and inflammation in population of some European regions, such as Norther Italy. In short, the exposure of air pollution, associated with Coronavirus infection, can be a driver of high rate of mortality in Italy (14.06%), Spain (11.90%), UK (14.37%), Belgium (16.40%), France (15.32%), etc. (cf., Center for System Science and Engineering at Johns Hopkins, 2020). The study by van Doremalen et al. (2020) revels that in China viral particles of SARS-CoV-2 may be suspended in the air for various minutes and this result can explain the high total number of infected people and deaths of COVID-19 infection in the USA, Spain, Russia, France, Italy, Brazil, Turkey, Iran, etc. (cf., Center for System Science and Engineering at Johns Hopkins, 2020). In general, these studies suggest the hypothesis that the atmosphere having a high level of air pollutant, associated with certain climatological factors, may support a longer permanence of viral particles in the air, fostering a diffusion of COVID-19 infection based on mechanisms of air pollution-to-human transmission in addition to human-to-human transmission (Frontera et a., 2020). In order to extend the investigation of these critical aspects in the development of COVID-19 outbreaks worldwide, in the presence of polluting industrialization, the goal of this study is to analyze the relation between infected people, air pollution, wind speed and inter-related renewable wind energy production that can explain some critical relationships determining the diffusion of COVID-19 and negative effect in environment and public health. This study has the potential to support long-run sustainable policy directed to foster a cleaner production for reducing and/or preventing the diffusion of future epidemics similar to COVID-19 infection.

**METHODS**

*1.1 Data sources and research setting*

This study focuses on fifty-five (*N*=55) cities that are provincial capitals in Italy, one of the countries with the highest number of deaths of COVID-19 infection: more than 31,360 units at 15 May, 2020 (cf., Lab24, 2020). Epidemiological data of COVID-19 infection are from Ministero della Salute (2020); data of polluting industrialization, air pollution and particulate compounds emissions are from Regional Agencies for Environmental

Protection in Italy (cf., Legambiente, 2019); climatological information are based on meteorological stations in Italian provinces (il Meteo, 2020); data of the density of population are from the Italian National Institute of Statistics (ISTAT, 2020) and finally, data concerning the production of wind energy per Italian regions are from Italian Transmission Operator called Terna (2020).

*1.2  Measurements*

- *Polluting industrialization and particulate compounds emissions.* Total days exceeding the limits set for $PM_{10}$ or for ozone in 2018 per Italian provincial capitals. Days of air pollution and particulate compounds emissions are a main factor that affects environment and public health. Moreover, 2018 as baseline year for air pollution and particulate compounds emissions data, it separates out the effects of COVID-19 infection.

- *Diffusion of COVID-19 infection.* Number of infected individuals on March-April, 2020

- *Climatological information.* Average wind speed km/h on February-March 2020

- *Indicators of interpersonal contact rates.* Population density of cities (individual / km²) in 2019

- *Production of renewable wind energy.* Power in MW of overall wind farms in all regions at January 2020

*1.3  Primary data analysis and statistics*

Descriptive statistics is performed categorizing Italian provincial capitals in groups, considering:

- *Renewable wind energy production*
  - cities with *high wind energy production* (seven regions in Italy have 94% of national production of wing energy)
  - cities with *low wind energy production* (regions that have 6% of national production of wing energy)
- *Polluting industrialization with and particulate compounds emissions*
  - Cities with *high polluting industrialization* (> 100 days per year exceeding the limits set for $PM_{10}$ or for ozone)
  - Cities with *low polluting industrialization* (≤ 100 days per year exceeding the limits set for $PM_{10}$ or for ozone)

Correlation and regression analyses verifies relationships between variables understudy. Regression analysis considers the number of infected people across Italian provincial capitals (variable *y*) as a linear function of the explanatory variable of total days exceeding the limits set for $PM_{10}$ (variable $x$).

The specification of linear relationship is a *log-log* model:

$$\log y_t = \alpha + \beta \log x_{t-1} + u \qquad [1]$$

$\alpha$ is a constant; $\beta$ = coefficient of regression; $u$ = error term

An alternative model [1] applies as explanatory variable the density of population per $km^2$ considering groups of cities with *high* or *low* level of polluting industrialization and particulate compounds emissions.

The estimation of equation [1] is also performed using a categorization of cities according to level of polluting industrialization and their location in regions with high and low intensity of wind energy production. Ordinary Least Squares (OLS) method is applied for estimating the unknown parameters of linear models [1]. Statistical analyses are performed with the Statistics Software SPSS® version 24.

## RESULTS

The wind energy production in Italy is in Table 1 per regions.

Table 1. Wind energy production in Italy per regions, January 2020

| Italian Regions | Number wind farms | Power [MW] |
|---|---|---|
| Abruzzo | 47 | 264.2 |
| Basilicata | 1413 | 1300.1 |
| Calabria | 418 | 1125.8 |
| Campania | 619 | 1734.6 |
| Emilia Romagna | 72 | 44.9 |
| Friuli Venezia Giulia | 5 | 0.0 |
| Lazio | 69 | 70.9 |
| Liguria | 33 | 56.8 |
| Lombardia | 10 | 0.0 |
| Marche | 51 | 19.2 |
| Molise | 79 | 375.9 |
| Piemonte | 18 | 23.8 |
| Puglia | 1176 | 2570.1 |
| Sardegna | 595 | 1105.3 |
| Sicilia | 884 | 1904.1 |
| Toscana | 126 | 143.0 |
| Trentino Alto Adige | 10 | 0.4 |
| Umbria | 25 | 2.1 |
| Valle d'Aosta | 5 | 2.6 |
| Veneto | 18 | 13.4 |

*Source*: Terna (2020)

Table 1 shows that seven regions in Italy (Molise, Puglia, Calabria, Basilicata, Campania, Sicilia and Sardegna) have the 94% of total wind energy production. These regions have at least over 1 GW of power, with the leadership of Puglia region (South-East Italy) having 2.5 GW. Italy had in January 2020 about 5,645 wind farms with almost 7,000 wind turbines of various power sizes. In particular, above 10 MW of power there are 313 plants for a total power of just over 9 GW (i.e., 9.07 GW). The most relevant power class ranges from 20 to 200 kW, with 3,956 systems having a total power of approximately 234 MW. As just mentioned, Puglia has the largest share of wind power installed in Italy: 24.8% of the total with 92 plants above 10 MW of power. Results of Italian province capitals, categorized in two groups belonging to regions with *high* or *low* wind energy production, suggest that cities in regions with a *high* production of wind energy (94% of total) have a very *low* number of infected individuals with COVID-19 infection (in March and April 2020), whereas cities located in regions with a low intensity of wind energy production (6% of total) have a very high number of infected individuals (Table 2).

Table 2. Descriptive statistics of Italian province capitals according to intensity of wind energy production

| *Cities in regions with 94% of wind energy production N=5* | Days exceeding limits set for $PM_{10}$ or ozone 2018 | Infected Individuals 17th March 2020 | Infected Individuals 7th April 2020 | Infected Individuals 27th April 2020 | Density inhabitants/km² 2019 | Wind km/h Feb-Mar 2020 |
|---|---|---|---|---|---|---|
| Mean | 48.00 | 59.80 | 505.60 | 708.20 | 2129.00 | 14.60 |
| Std. Deviation | 30.27 | 90.84 | 646.12 | 949.19 | 3384.10 | 5.45 |
| *Cities in regions with 6% of wind energy production N=50* | | | | | | |
| Mean | 79.44 | 475.58 | 2119.68 | 3067.67 | 1385.76 | 8.10 |
| Std. Deviation | 41.70 | 731.11 | 2450.71 | 3406.67 | 1489.31 | 3.08 |

Table 2 also shows that cities in regions with low production of wind energy (6% of total) have a higher level of polluting industrialization than cities with a high production of wind energy (about 70 polluted days *vs*. 48 polluted days exceeding $PM_{10}$ or ozone per year). This preliminary result suggests that regions with a high intensity of wind-based renewable energy and low polluting industrialization have also a low diffusion of COVID-19 infection in society. In order to confirm this result, table 3 considers polluting industrialization of cities: especially, cities with high polluting industrialization and particulate compounds emissions (>100 days exceeding limits set for $PM_{10}$ or ozone per year) and low production of wind energy, they have a very high level of infected individuals in March and April 2020, in an environment with high average density of population and low average intensity of wind speed.

Table 3.  Descriptive statistics of Italian provincial capitals according to polluting industrialization and particulate compounds emissions

| *Cities with high polluting industrialization:* >100days exceeding limits set for $PM_{10}$ N=20 | Days exceeding limits set for $PM_{10}$ or ozone 2018 | Infected Individuals 17th March 2020 | Infected Individuals 7th April 2020 | Infected Individuals 27th April 2020 | Density inhabitants/km² 2019 | Wind km/h Feb-Mar 2020 |
|---|---|---|---|---|---|---|
| Mean | 125.25 | 881.70 | 3650.00 | 4838.05 | 1981.40 | 7.67 |
| Std. Deviation | 13.40 | 1010.97 | 3238.82 | 4549.41 | 1988.67 | 2.86 |
| *Cities with low polluting industrialization:* <100days exceeding limits set for $PM_{10}$ N=35 | | | | | | |
| Mean | 48.77 | 184.11 | 1014.63 | 1637.21 | 1151.57 | 9.28 |
| Std. Deviation | 21.37 | 202.76 | 768.91 | 1292.26 | 1466.28 | 4.15 |

Table 4.  Correlation

| | *Cities in regions with 94% of wind energy production* | *Cities in regions with 6% of wind energy production* |
|---|---|---|
| | *Log* Days exceeding limits set for $PM_{10}$ or ozone 2018 | *Log* Days exceeding limits set for $PM_{10}$ or ozone 2018 |
| *Log* Infected Individuals 17th March, 2020 | | |
| Pearson Correlation | .81 | .69** |
| *Log* Infected individuals 7th April, 2020 | | |
| Pearson Correlation | .74 | .55** |
| *Log* Infected individuals 27th April, 2020 | | |
| Pearson Correlation | .69 | .36** |

Note:  **. Correlation is significant at the 0.01 level (2-tailed)

Table 4 shows that cities of regions with less than 6% of wind energy production, they have a high positive correlation between polluting industrialization and infected individuals of COVID-19 infection at 17th March (*r*=.69, *p*-value<.01), 7th April (*r*=.55, *p*-value<.01) and 27th April, 2020 (*r*=.36, *p*-value<.01). In regions with a high intensity of wind production, results are not significant.

Table 5.    Parametric estimates of the relationship of *Log* Infected individuals on *Log* polluting industrialization considering the groups of cities in regions with *high* or *low* production of wind energy

|  | Cities in regions with 94% of wind energy production |  | Cities in regions with 6% of wind energy production |
|---|---|---|---|
|  | Explanatory variable: *Log* Days exceeding limits set for $PM_{10}$ or ozone |  | Explanatory variable: *Log* Days exceeding limits set for $PM_{10}$ or ozone |
| ↓DEPENDENT VARIABLE | 2018 | ↓DEPENDENT VARIABLE | 2018 |
| ***log* infected** |  | ***log* infected** |  |
| **7th April, 2020** |  | **7th April, 2020** |  |
| Constant $\alpha$ | .70 | Constant $\alpha$ | 3.39*** |
| (St. Err.) | (2.64) | (St. Err.) | (.85) |
| Coefficient $\beta 1$ | 1.34 | Coefficient $\beta 1$ | .92*** |
| (St. Err.) | (.70) | (St. Err.) | (.20) |
| R² (St. Err. of Estimate) | .55 (.86) | R² (St. Err. of Estimate) | .31 (.82) |
| F | 3.65 | F | 21.28*** |

*Note*: Explanatory variable: *log* Days exceeding limits set for $PM_{10}$ or ozone 2018; dependent variable *log* infected individuals
 \*\*\*   *p*-value<0.001

Table 5 suggests that polluting industrialization, in areas with low production of wind energy, explains the number of infected individuals of COVID-19. In particular,

o cities in regions with 94% of wind energy production have not significant results because of low number of cases in sample

o instead, in cities of regions with 6% of wind energy production, an increase of 1% of polluting industrialization, measured with days exceeding limits set for $PM_{10}$, it increases the expected number of infected by about 0.92% (P<.001).

*Figure 1 shows a visual representation of regression lines: cities having a higher production of renewable energy tend to have a low number of total infected individuals driven by polluting industrialization.*

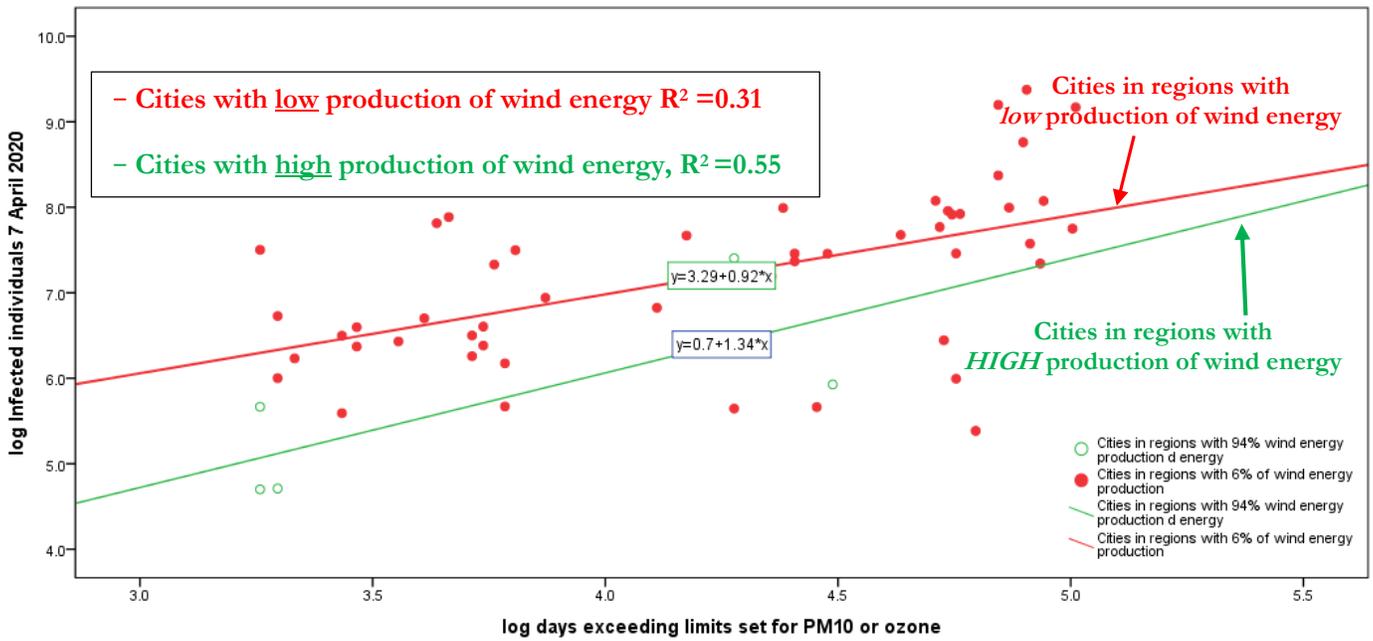

**Figure 1–** Regression lines of *Log* Infected Individuals on *Log* polluting industrialization according to production of wind energy of cities. Note: *This result suggests that diffusion of COVID-19 infection increases with polluting industrialization in regions having low production of wind energy, i.e., with a less sustainable production.*

In order to confirm this findings, table 6 considers cities with a high and low polluting industrialization.

Table 6.    Parametric estimates of the relationship of *Log* Infected individuals on *Log* Density inhabitants/km² 2019, considering the groups of cities with *high* and *low* polluting industrialization

|  | Cities with *low* polluting industrialization | | Cities with *high* polluting industrialization |
|---|---|---|---|
|  | *Explanatory variable:* *Log* Density inhabitants/km² 2019 |  | *Explanatory variable:* *Log* Density inhabitants/km² 2019 |
| ↓Dependent variable |  | ↓Dependent variable |  |
| *log* infected 7th April, 2020 |  | *log* infected 7th April, 2020 |  |
| Constant α | 4.976 | Constant α | 1.670 |
| (St. Err.) | (.786) | (St. Err.) | (1.491) |
| Coefficient β 1 | .252* | Coefficient β 1 | .849*** |
| (St. Err.) | (.120) | (St. Err.) | (.205) |
| R² (St. Err. of Estimate) | .119 | R² (St. Err. of Estimate) | .488 |
| F | 17.168*** | F | 4.457* |

*Note*: Explanatory variable: *log* Density inhabitants/km² in 2019; dependent variable *log* infected individuals
***      $p$-value<0.001
**       $p$-value<0.01
\*        $p$-value<0.05

Table 6 reveal that *in cities with low polluting industrialization and low particulate compounds emissions,* an increase of 1% of the density of population, it increases the expected number of infected individuals by about 0.25% (*P*=.042); whereas, *in cities with high polluting industrialization,* an increase of 1% of the density of population, it increases the expected number of infected individuals by about 85% (*P*<.001). Figure 2 shows regression lines on 7th April 2020, in the middle phase of COVID-19 outbreak in Italy: regions with a polluting industrialization generating an atmosphere rich of air pollutants that associated with a climate factor of low wind speed support a stronger of diffusion of COVID-19 infection.

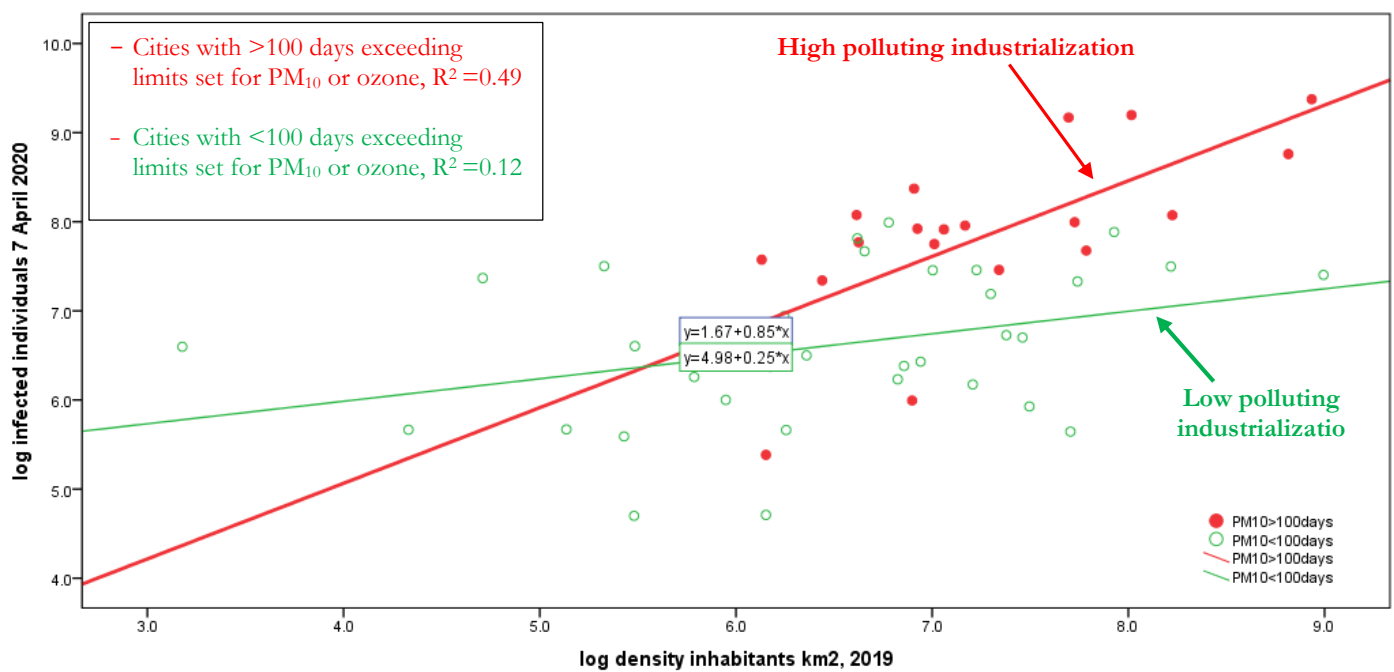

**Figure 2**: Regression line of *Log* Infected people on *Log* population density inhabitants, considering the groups of cities with *high* or *low* polluting industrialization
Note: *This result reveals that diffusion of COVID-19 is higher in cities with high polluting industrialization*

In addition, if we consider regions with high/low air pollution and particulate compounds emissions, using arithmetic mean of days exceeding limits set for $PM_{10}$ or ozone of cities, the percentage of infected individuals and total deaths, weighted with population of these regions, reveals that about 74.50% of infected individuals and about 81% of total death in Italy because of COVID-19 infection are in regions with high air pollution and polluting industrialization and with low production of the renewable energy based on wind resource.

# DISCUSSION AND LIMITATIONS

This new study finds that geo-environmental factors may have accelerated the spread of COVID-19 in northern Italian cities, leading to a higher number of infected individuals and deaths. This study analyzed data on COVID-19 cases alongside environmental and wind energy data. It found that cities with little wind and frequently high levels of air pollution — exceeding safe levels of ozone or particulate matter — had higher numbers of COVID-19 related infected individuals and deaths. These findings suggest that the current pandemic of Coronavirus disease and future epidemics similar to COVID-19 cannot be solved *only* with research and practice in medicine, immunology and microbiology but *also* with the development of industrial instruments directed to a sustainable and cleaner production (Coccia, 2019). These findings here provide valuable insight into geo-environmental and industrial factors that may accelerate the diffusion of COVID-19 and similar viral agents. The main results of the study, based on case study of COVID-19 outbreak in Italy, are:

o   The diffusion of COVID-19 in Italy has a high association with high polluting industrialization in cities

o   Cities having a high production of wind energy, associated with low polluting industrialization, have a low diffusion of COVID-19 infection and a lower number of total deaths.

Considering the results just mentioned, the question is:

*what is the link between diffusion of COVID-19 infection, polluting industrialization and renewable wind energy in specific regions?*

Results suggest that, among Italian provincial capitals, the number of infected people is higher in cities with polluting industrialization, cities located in hinterland zones (i.e. away from the coast), cities having a low average intensity of wind speed and cities with a lower temperature. In hinterland cities (mostly those bordering large urban conurbations, such as Bergamo, Brescia, Lodi, close to Milan in Lombardy region of North West Italy etc.) with a high polluting industrialization, coupled with low wind speed and wind energy production, the average number of infected people in April 2020 more than doubled that of more windy cities with renewable energy production. Therefore, cities in regions with a high production of wind energy in Italy, they also have a low polluting industrialization, low air pollution and particulate compounds emissions., in an environment with a high intensity of wind speed that sustains clean days from air pollution, particulate compounds emissions that current studies suggest the higher diffusion of Coronavirus infection (Fattorini and Regoli, 2020). As a matter of fact, cities with high polluting industrialization, mainly in Northern Italy (also having a low wind speed and as a consequence low wind energy production), have a stagnation of air pollution in the atmosphere that can support diffusion of COVID-19 infection (Contini and Costabile, 2020; Conticini et al., 2020). The implications for a sustainable policy are clear: *COVID-19 outbreak has low diffusion in regions with low polluting industrialization and high production of renewable (wind) energy.* Northern Italian region covered by the study, as a consequence, in future should reduce pollution industrialization and particulate compounds emissions, so that the accelerated transmission dynamics of COVID-19 viral infectivity are not triggered. In order to reinforce these conclusions with a perspective of sustainable policies, Xu et al. (2020) found out the effect of moisture on explosive growth in fine particulate matter (PM), and propose a new approach for the simulation of fine PM growth and dissipation in ambient air. In particular, winds significantly aid the dissipation of fine PM, and high concentrations of fine PM only persisted for a very short time and dissipated after several hours. The role of climatological factors, such as wind speed and direction, temperature, and humidity are critical for urban ventilation and the pollutant concentration in the streets of cities (Yuan et al. 2019). Considering the benefit of wind as resource that can reduce air pollution and as a consequence viral infectivity with main public health benefits, Gu et al. (2020) argue that a strategy to enhance air quality in cities is improving urban ventilation: the ability of an urban area to dilute pollutants and heat by improving the exchange of air between areas within and above the urban canopy. Of course, urban ventilation is a function of a manifold urban geometry characteristics, e.g., frontal and plan area density, and the aspect ratio of urban morphology. Studies show that variations of building height have beneficial effects in terms of breathability levels, whereas larger aspect ratios of urban canyons can lead to high levels of pollutant concentrations inside the streets of cities. Hence, cities located in hinterland zones of the Northern Italian region with low wind speed have an urban climatology and aspects of urban and regional topography that sustain the stagnation of air pollution that can support the spread of viral infectivity in fall and winter season. Hence, these regions have to reduce the level of particulate compounds emissions applying long-run sustainable polices directed to reduce polluting industrialization and support the production of renewable energy (Wang and Zhu, 2020). In fact, health and economic benefits associated with national and local reduction of air pollution are now rarely contested. Cui et al. (2020), based on a study in China, show that where reductions in ambient air pollution and particulate compounds emissions have avoided more than 2,300 premature deaths and more than 15,80 related morbidity cases in 2017, with a total of about US$ 318 million in economic benefits. In addition, these scholars argue that reduction of $PM_{2.5}$ concentrations to 15 µg/m$^3$ would result in reductions of 70% in total $PM_{2.5}$-related non-accidental mortality and 95% in total $PM_{2.5}$-related morbidity, with economic benefits of more than US$ 1,289.5 million. In short, sustainable policies that reduce air pollution and particulate compounds emissions generate significant environmental, public health, social and economic benefits. This study suggests that in order to prevent epidemics similar to COVID-19 and other infections, nations have to apply a sustainable policy directed to reduce air pollution that affects public health and amplifies the negative effects of airborne viral diseases. In addition, the policy for a sustainable development has to consider the urban climatology with the study of climatic properties of urban areas (Gu et al., 2020) and support renewable energy, such as wind resource, that create the environmental conditions for the reduction of air pollution on trans-regional level (Wang and Zhu, 2020). **Moreover, high surveillance and proper biosafety procedures in public and private institutes of virology that study viruses and new viruses to avoid that may be accidentally spread in surrounding environments with damages for population and vegetation. In this context, international collaboration among scientists is basic to address these risks, support decisions of policymakers to prevent future pandemic creating potential huge socioeconomic issues worldwide (cf., Coccia** and Wang, 2016; Coccia, 2020a)[1]. In fact, following the COVID-19 outbreak, The Economist Intelligence Unit (EIU) points out that the global economy may contract of about by 2.2% and Italy by -7% of real GDP growth % in 2020 (EIU, 2020). Italy and other advanced countries should introduce organizational, product and process innovations to cope with future viral threats, such as the expansion of hospital capacity and testing capabilities, to reduce diagnostic and health system delays also using artificial intelligence, and as a consequence new ICT technologies for alleviating and/or eliminating effective interactions between infectious and susceptible individuals, and finally of course to develop effective vaccines and antivirals that can counteract future global public health threat in the presence of new epidemics similar to COVID-19 (Chen et al, 2020; Wilder-Smith et al 2020; Riou and Althaus, 2020; Yao et al., 2020; cf., Coccia, 2005, 2009, 2015a, 2017b, 2018, 2019a, 2020b; Coccia and Watts, 2020). In short, the concentration in specific areas of a combination of climate with low wind, a specific urban climatology of hinterland cities, high polluting industrialization, aspects of regional topography and physical geography sustains, in fall and winter season, the stagnation of air pollution and particulate compounds emissions that seems to have supported the spread of COVID-19 infection (cf., Contini and Costabile, 2020; Conticini et al., 2020; Fattorini and Regoli, 2020). New findings here show that geo-environmental factors may have accelerated the spread of COVID-19 in northern Italian cities, leading to a higher number of infected individuals and deaths.

The results here also suggested that, among Italian provincial capitals, the number of infected people was higher in cities with >100 days per year exceeding limits set for PM10 or ozone, cities located in hinterland zones (i.e. away from the coast), cities having a low average intensity of wind energy production and cities with a lower temperature. In hinterland cities (mostly those bordering large urban conurbations) with a high number of days exceeding PM10 and ozone limits, coupled with low wind speed, the average number of infected people in April more than doubled.

---

[1] Socioeconomic shocks can lead to a general increase of prices, high public debts, high unemployment, income inequality and as a consequence violent behavior (Coccia, 2016, 2017, 2017a).

These findings provide valuable insight into geo-environmental and industrial factors that may accelerate the diffusion of COVID-19 and similar viral agents. In this context, a proactive strategy to help cope with future epidemics should concentrate on reducing levels of air pollution in hinterland and polluted cities.

However, these conclusions are of course tentative because there are several challenges to such studies, particularly in real time because the sources can only capture certain aspects of the on-going complex relations between polluting industrialization, diffusion of viral infectivity and other resources of economic systems. This study therefore encourages further investigations on these aspects of the diffusion of COVID-19 outbreaks in highly industrialized areas to design appropriate sustainable policies that can provide lung-run public health measures to reduce air pollution and control the spread of infection similar to COVID-19 (Ou et al., 2020). Overall, then, in the presence of high polluting industrialization and low renewable energy production of regions that can support diffusion of epidemics in environment with high level of air pollution and particulate compounds emissions, this study has to suggest that a comprehensive strategy to prevent future epidemics similar to COVID-19 must be designed in terms of sustainability science with a high incidence of cleaner production in socioeconomic systems.

**Declaration of competing interest**

The author declares that he has no known competing financial interests or personal relationships that could have appeared to influence the work reported in this paper. No funding was received for this study.

**REFERENCES**

Center for System Science and Engineering at Johns Hopkins 2020. Coronavirus COVID-19 Global Cases, https://gisanddata.maps.arcgis.com/apps/opsdashboard/index.html#/bda7594740fd40299423467b48e9ecf6 (accessed in 9th May 2020).

Chen S., Yang J., Yang W., Wang C, Barnighausen T. 2020. COVID-19 control in China during mass population movements at New Year. Lancet; 395: 764–66.

Coccia M. 2005. Metrics to measure the technology transfer absorption: analysis of the relationship between institutes and adopters in northern Italy. International Journal of Technology Transfer and Commercialization, vol. 4, n. 4, pp. 462-486. https://doi.org/10.1504/IJTTC.2005.006699

Coccia M. 2009. A new approach for measuring and analyzing patterns of regional economic growth: empirical analysis in Italy, Italian Journal of Regional Science- Scienze Regionali, vol. 8, n. 2, pp. 71-95. DOI: 10.3280/SCRE2009-002004

Coccia M. 2014. Path-breaking target therapies for lung cancer and a far-sighted health policy to support clinical and cost effectiveness, Health Policy and Technology, vol. 1, n. 3, pp. 74-82. https://doi.org/10.1016/j.hlpt.2013.09.007

Coccia M. 2015. The Nexus between technological performances of countries and incidence of cancers in society, Technology in Society, vol. 42, August, pp. 61-70. DOI: http://doi.org/10.1016/j.techsoc.2015.02.003

Coccia M. 2015a. Technological paradigms and trajectories as determinants of the R&D corporate change in drug discovery industry. Int. J. Knowledge and Learning, vol. 10, n. 1, pp. 29–43. http://dx.doi.org/10.1504/IJKL.2015.071052

Coccia M. 2016. The relation between price setting in markets and asymmetries of systems of measurement of goods, The Journal of Economic Asymmetries, vol. 14, part B, November, pp. 168-178, https://doi.org/10.1016/j.jeca.2016.06.001

Coccia M. 2017. A Theory of general causes of violent crime: Homicides. Income inequality and deficiencies of the heat hypothesis and of the model of CLASH, Aggression and Violent Behavior, vol. 37, November-December, pp. 190-200, https://doi.org/10.1016/j.avb.2017.10.005

Coccia M. 2017a. Asymmetric paths of public debts and of general government deficits across countries within and outside the European monetary unification and economic policy of debt dissolution, The Journal of Economic Asymmetries, vol. 15, June, pp. 17-31, https://doi.org/10.1016/j.techfore.2010.02.003

Coccia M. 2017b. The Fishbone diagram to identify, systematize and analyze the sources of general purpose technologies. Journal of Social and Administrative Sciences, J. Adm. Soc. Sci. – JSAS, vol. 4, n. 4, pp. 291-303, http://dx.doi.org/10.1453/jsas.v4i4.1518

Coccia M. 2018. Theorem of not independence of any technological innovation, Journal of Economics Bibliography, vol. 5, n. 1, pp. 29-35, http://dx.doi.org/10.1453/jeb.v5i1.1578

Coccia M. 2019. Theories of Development. A. Farazmand (ed.), Global Encyclopedia of Public Administration, Public Policy, and Governance, Springer Nature Switzerland AG, ISBN: 978-3-319-20927-2, https://doi.org/10.1007/978-3-319-31816-5_939-1

Coccia M. 2019a. Why do nations produce science advances and new technology? Technology in society, vol. 59, November, 101124, pp. 1-9, https://doi.org/10.1016/j.techsoc.2019.03.007

Coccia M. 2020. Factors determining the diffusion of COVID-19 and suggested strategy to prevent future accelerated viral infectivity similar to COVID, Science of the Total Environment, volume, 729, Article Number: 138474, https://doi.org/10.1016/j.scitotenv.2020.138474

Coccia M. 2020a. The evolution of scientific disciplines in applied sciences: dynamics and empirical properties of experimental physics, Scientometrics, pp. 1-37 https://doi.org/10.1007/s11192-020-03464-y

Coccia M. 2020b. Deep learning technology for improving cancer care in society: New directions in cancer imaging driven by artificial intelligence. Technology in Society, vol. 60, February, pp. 1-11, https://doi.org/10.1016/j.techsoc.2019.101198

Coccia M., Wang L. 2015. Path-breaking directions of nanotechnology-based chemotherapy and molecular cancer therapy, Technological Forecasting & Social Change, 94(May):155–169. https://doi.org/10.1016/j.techfore.2014.09.007

Coccia M., Wang L. 2016. Evolution and convergence of the patterns of international scientific collaboration, Proceedings of the National Academy of Sciences of the United States of America, vol. 113, n. 8, pp. 2057-2061, www.pnas.org/cgi/doi/10.1073/pnas.1510820113.

Coccia M., Watts J. 2020. A theory of the evolution of technology: technological parasitism and the implications for innovation management, Journal of Engineering and Technology Management, vol. 55 (2020) 101552, S0923-4748(18)30421-1,https://doi.org/10.1016/j.jengtecman.2019.11.003

Conticini E., Frediani B., Caro D. 2020. Can atmospheric pollution be considered a co-factor in extremely high level of SARS-CoV-2 lethality in Northern Italy? Environmental Pollution, Volume 261,114465, https://doi.org/10.1016/j.envpol.2020.114465.

Contini, D.; Costabile, F. 2020. Does Air Pollution Influence COVID-19 Outbreaks? Atmosphere, 11, 377.

Cui L., Zhou J., Peng X., Ruan S., Zhang Y. 2020. Analyses of air pollution control measures and co-benefits in the heavily air-polluted Jinan city of China, 2013-2017.Sci Rep. 2020 Mar 25;10(1):5423. doi: 10.1038/s41598-020-62475-0.

Dantas Guilherme, Siciliano Bruno, Bruno Boscaro França, Cleyton M. da Silva, Graciela Arbilla, 2020. The impact of COVID-19 partial lockdown on the air quality of the city of Rio de Janeiro, Brazil, Science of The Total Environment, Volume 729, 139085, https://doi.org/10.1016/j.scitotenv.2020.139085.

EIU 2020. COVID-19 to send almost all G20 countries into a recession, 26th Mar 2020.

Fattorini D., Regoli F. 2020. Role of the chronic air pollution levels in the Covid-19 outbreak risk in Italy, Environmental Pollution, Volume 264,2020,114732, https://doi.org/10.1016/j.envpol.2020.114732.

Frontera A., Claire Martin, Kostantinos Vlachos, Giovanni Sgubin, 2020. Regional air pollution persistence links to COVID-19 infection zoning, J Infect. 2020 Apr 10 doi: 10.1016/j.jinf.2020.03.045

Gu K., Yunhao Fang, Zhao Qian, Zhen Sun & Ai Wang 2020. Spatial planning for urban ventilation corridors by urban climatology, Ecosystem Health and Sustainability, 6:1, 1747946, DOI: 10.1080/20964129.2020.1747946

Il meteo 2020. Medie e totali mensili. https://www.ilmeteo.it/portale/medie-climatiche (Accessed March 2020).

ISTAT 2020. The Italian National Institute of Statistics-Popolazione residente al 1 gennaio, http://dati.istat.it/Index.aspx?DataSetCode=DCIS_POPRES1

Lab24 2020. Coronavirus in Italia, i dati e la mappa. Il Sole24ORE. https://lab24.ilsole24ore.com/coronavirus/ (Accessed, 9 May, 2020)

Legambiente 2019. Mal'aria 2019, il rapporto annuale sull'inquinamento atmosferico nelle città italiane. https://www.legambiente.it/malaria-2019-il-rapporto-annuale-annuale-sullinquinamento-atmosferico-nelle-citta-italiane/ (ACCESSED March 2020)

Lewtas J. 2007. Air pollution combustion emissions: Characterization of causative agents and mechanisms associated with cancer, reproductive, and cardiovascular effects. Mutation Research, vol. 636, Issues 1–3, pp. 95-133, https://doi.org/10.1016/j.mrrev.2007.08.003.

Ministero della Salute 2020. Covid-19 - Situazione in Italia. http://www.salute.gov.it/portale/nuovocoronavirus/dettaglioContenutiNuovoCoronavirus.jsp?lingua=italiano&id=5351&area=nuovoCoronavirus&menu=vuoto (Accessed April 2020)

Ogen Yaron, 2020. Assessing nitrogen dioxide (NO2) levels as a contributing factor to coronavirus (COVID-19) fatality, Science of The Total Environment, Volume 726,2020,138605, https://doi.org/10.1016/j.scitotenv.2020.138605.

Ou Y., J. West, S. Smith, Chris Nolte, Dan Loughlin 2020. Air Pollution Control Strategies Directly Limiting National Health Damages in the U.S. Nature Communications, 11:957, https://doi.org/10.1038/s41467-020-14783-2

Riou J., Althaus C.L. 2020. Pattern of early human-to-human transmission of Wuhan 2019 novel coronavirus (2019-nCoV), December 2019 to January 2020. Euro Surveill 2020; 25: 2000058.

Saadat S., Deepak Rawtani, Chaudhery Mustansar Hussain 2020. Environmental perspective of COVID-19, Science of The Total Environment, Volume 728,2020,138870, https://doi.org/10.1016/j.scitotenv.2020.138870.

Terna 2020. Fonte rinnovabili. Wind Energy. https://www.terna.it/it/sistema-elettrico/dispacciamento/fonti-rinnovabili (Accessed May, 2020)

van Doremalen N., Bushmaker T., Morris D.H., Holbrook M.G., Gamble A., Williamson B.N., Tamin A., Harcourt J.L., Thornburg N.J., Gerber S.I., Lloyd-Smith J.O., de Wit E., Munster V.J. 2020.Aerosol and Surface Stability of SARS-CoV-2 as Compared with SARS-CoV-1. N Engl J Med. Apr 16;382(16):1564-1567. doi: 10.1056/NEJMc2004973.

Wang Z., Zhu Yongfeng 2020. Do energy technology innovations contribute to CO2 emissions abatement? A spatial perspective, Science of The Total Environment, Volume 726,2020,138574, https://doi.org/10.1016/j.scitotenv.2020.138574.

Wilder-Smith A., Chiew C.J., Lee V.J. 2020. Can we contain the COVID-19 outbreak with the same measures as for SARS? Lancet Infect Dis 2020; published online March 5. https://doi.org/10.1016/S1473-3099(20)30129-8.

Xu J., Fahua Zhu, Sheng Wang, Xiuyong Zhao, Ming Zhang, Xinlei Ge, Junfeng Wang, Wenxin Tian, Liwen Wang, Liu Yang, Li Ding, Xiaobo Lu, Xinxin Chen, Youfei Zheng, Zhaobing Guo 2020. A preliminary study on wind tunnel simulations of the explosive growth and dissipation of fine particulate matter in ambient air, Atmospheric Research, Volume 235,2020,104635, https://doi.org/10.1016/j.atmosres.2019.104635.

Yao X., Ye F., Zhang M., et al. In vitro antiviral activity and projection of optimized dosing design of hydroxychloroquine for the treatment of severe acute respiratory syndrome coronavirus 2 (SARS-CoV-2). Clin Infect Dis 2020; published online March 9. DOI:10.1093/cid/ciaa237.

Yuan M., Y. Song, Y. Huang, H. Shen, and T. Li. 2019. Exploring the Association between the Built Environment and Remotely Sensed PM2.5 Concentrations in Urban Areas. Journal of Cleaner Production 220: 1014–1023. doi:10.1016/j.jclepro.2019.02.236